\documentclass[aps,prc,twocolumn,groupedaddress,showpacs]{revtex4}
\usepackage{graphicx}

\begin{document}


\title{Determination of S$_{17}$(0) from published data}
\author{R. H. Cyburt}
\author{B. Davids}
\author{B. K. Jennings}
\affiliation{TRIUMF, 4004 Wesbrook Mall, Vancouver BC V6T 2A3, Canada}


\date{\today}

\begin{abstract}
The experimental landscape for the $^7$Be$+p$ radiative capture
reaction is rapidly changing as new high precision data become
available. We present an evaluation of existing data, detailing the
treatment of systematic errors and discrepancies, and show how they
constrain the astrophysical $S$ factor ($S_{17}$), independent of any
nuclear structure model. With theoretical models robustly determining
the behavior of the sub-threshold pole, the extrapolation error can be
reduced and a constraint placed on the slope of $S_{17}$. Using only radiative capture data, we find
$S_{17}(0)=20.7\pm0.6(stat)\pm1.0(syst)$ eV~b if data sets are
completely independent, while if data sets are
completely correlated we find $S_{17}(0)=21.4\pm0.5(stat)\pm1.4(syst)$
eV~b. The truth likely lies somewhere in between these two limits. Although we employ a formalism capable of treating discrepant data, we note that the central value of the $S$ factor is dominated by the recent high precision data of Junghans \emph{et al.}, which imply a substantially higher value than other radiative capture and indirect measurements. Therefore we conclude that further progress will require new high precision data with a detailed error budget.\end{abstract}


\maketitle

\section{Introduction}

The comparison of measured and predicted $^{8}$B solar neutrino fluxes
represents a test of solar models and an opportunity to learn about
the properties of the electron neutrinos produced in the Sun. Recent
measurements at SNO \cite{ahmed04} have determined the total flux of
active neutrinos emitted in the $\beta^+$ decay of $^8$B with a
combined statistical and systematic precision of 9\%. The implications
of solar and reactor neutrino flux measurements for neutrino mixing
parameters are explored in, e.g.,
\cite{maltoni03,bandyopadhyay04}. Theoretical predictions of the
$^{8}$B solar neutrino flux are now substantially more uncertain than
the experimental measurements. Recent predictions have uncertainties
of 20\% \cite{watanabe01} and 15\% \cite{couvidat03}. The latest
effort to estimate the theoretical uncertainty found a value of 23\%
\cite{bahcall04}. This error is completely dominated by the
uncertainty in the heavy element abundance of the Sun, which has
recently been revised to a level 2.5 times larger than the previous
adopted value \cite{bahcall01}. The contribution to the theoretical
error budget made by S$_{17}$ has now been estimated at 3.6\% based on
the recommendation of S$_{17}$(0)~=~21.4~$\pm$~0.5~(expt.)~$\pm$~0.6~(theor.)~eV~b given in \cite{junghans03}. Even if the uncertainty
due to S$_{17}$ represents a small fraction of the total theoretical
uncertainty, in our view an independent determination of its value
based on the available experimental data and theoretical models is
worthwhile. Here we provide reliable determinations of the total
uncertainty in the $S$ factor, and find that nearly all previous
analyses have underestimated the error.

In this paper we detail our fitting procedure, presenting
formalisms for propagating systematic errors and combining multiple
data sets, some of which were first presented in~\cite{cyburt04}. We then
describe the available data relevant for $^7$Be(p,$\gamma$)$^8$B, and
present the data sets used in this analysis. Next we briefly discuss the structure models used
to extrapolate experimental data to solar energies, and the question
of how the models can be tested. Finally, we present our constraints
on the astrophysical $S$ factor using: (1) the energy dependence of the best structure models, (2) a pole model parametrization independent of structure models, and (3) a
constrained pole model parametrization, where theory is used to
robustly determine the behavior of the subthreshold pole.

\section{Fitting Procedure}\label{fit}

It is often desirable to determine an average or best-fit representation of experimental data, and the
uncertainties in such a representation. The standard techniques are generally not discussed in the literature, and are assumed to be well known. We detail here a formalism needed to properly take into account correlated data and systematic errors.

When modeling data, one generally uses a maximum likelihood or minimum
$\chi^2$ formalism to determine the best fit. The $\chi^2$ is usually
defined as:
\begin{equation}
\label{eqn:oldchi2}
\chi^2 = \sum_i \left( \frac{y(x_i)-\mu_i}{\sigma_i} \right)^2,
\end{equation}
where $\mu_i$, and $\sigma_i$ are the mean and standard deviation of
the $i^{th}$ data point and $y(x)$ is the value calculated from the model we are using to
describe the data as a function of the independent variable $x$
(e.g. $y(x)\rightarrow S_{17}(E)$). As an example, a linear model
consists of $P$ parameters and $P$ basis functions (i.e. $y(x) =
\sum_{p=1}^P a_p {\rm Y}_p(x)$). Minimizing $\chi^2$ defined in this way also minimizes the variance in the best fit model, but is
limited in that it assumes the data points are independent
and there is no explicit prescription for dealing with systematic errors or
multiple data sets. A more general treatment has been described recently in an
analysis of reaction cross sections relevant for big bang
nucleosynthesis~\cite{cyburt04}. We shall adopt a similar procedure for
studying the $^7{\rm Be}({\rm p},\gamma)^8{\rm B}$ reaction. A clear understanding of how statistical and
systematic uncertainties propagate through the analysis is an essential aspect of this approach. It
is assumed that the dominant systematic error is the normalization
uncertainty, which can be parametrized by a relative uncertainty,
$\epsilon_n$, for each data set $n$. This global normalization error
induces correlations between data points in the same data set. The
correlation between two data points is given by~\cite{cyburt04}:
\begin{equation}
\label{eqn:datcov}
C_{i_n,j_n} = (1 + \epsilon_n^2)\sigma_{i_n}^2\delta_{i_n,j_n} +
\epsilon_n^2\mu_{i_n}\mu_{j_n},
\end{equation}
where $\sigma_{i_n}$ is now the statistical uncertainty of the
$i^{th}$ data point of data set n and $\delta_{i_n,j_n}$ is the Kronecker delta.
Using this correlation matrix, one can define a more general $\chi^2$:
\begin{equation}
\label{newchi2}
\chi^2 = \sum_{i,j} C_{i,j}^{-1}[y(x_i)-\mu_i][y(x_j)-\mu_j],
\end{equation}
where we have left off the data set subscript $n$ for clarity. Given the simple form of the covariance matrix in Eq.~\ref{eqn:datcov}, the
inverse can be found analytically~\cite{cyburt04}, and we obtain
\begin{equation}
\label{eqn:invdatcov}
C_{i,j}^{-1} = \frac{\delta_{i,j}}{(1 +
\epsilon^2)\sigma_{i}^2} -
\frac{\frac{\epsilon^2\mu_{i}\mu_{j}}{(1 +
\epsilon^2)^2\sigma_{i}^2\sigma_{j}^2}}{1 + \frac{\epsilon^2}{(1
+ \epsilon^2)}\sum_{k} \left(
\frac{\mu_{k}}{\sigma_{k}}\right)^2},
\end{equation} where we have again suppressed the index $n$.
When the systematic errors are smaller than the statistical errors
($\epsilon_n\mu_{i_n} < \sigma_{i_n}$) or in the limit of large
data sets, the $\chi^2$ of Eq.~\ref{newchi2} reduces to that of
Eq.~\ref{eqn:oldchi2}, with the statistical error weighting the
$\chi^2$ rather than the total error. 

We stress that the $\chi^2$ quantity is a
statistical device only; its minimum value tells us only how good the
fit is within statistical uncertainties and its curvature tells us only about
the statistical uncertainties in the fit. This is seen as a
$1/\sqrt{N}$ scaling in the best fit uncertainty, where $N$ is the
number of data points. Systematic errors are not reduced by this
factor. The total uncertainty then consists of the statistical error
and the intrinsic normalization error. Statistical errors contain
no information about the quality of the fit. This can be seen by
arbitrarily shifting data points away from each other, keeping their
absolute uncertainties the same. We need an additional measure of how far data
fall from the best fit. How then do we address the quality of the
fit?

In Ref.\ \cite{cyburt04}, such a discrepancy error is defined as a measure of the
fit quality. The discrepancy error is the weighted dispersion of the
data relative to the best fit: 
\begin{equation}
\label{eqn:disc}
\epsilon_{disc}^2 = \frac{\sum_{i,j} {\mathcal C}_{i,j}^{-1} [y(x_i)-\mu_i][y(x_j)-\mu_j]}{\sum_{i,j} {\mathcal C}_{i,j}^{-1} y(x_i)y(x_j)}.
\end{equation}
The absolute size of this discrepancy error tells us how well the data
are described by the best fit, while its size relative to the
intrinsic normalization error quantifies possible unknown
systematics. One may be tempted to reduce the size of the
discrepancy error by the number of degrees of freedom
(e.g. $\chi^2\rightarrow \chi^2/\nu$), but this is inappropriate because it
assumes that the unknown errors we are trying to take into account can
be propagated through the data analysis. As discussed earlier,
systematics are not reduced by $1/\sqrt{N}$ as are statistical
errors. We adopt a total normalization
error defined as the quadrature sum of the intrinsic normalization
error and this discrepancy error, as was done in Ref.\ \cite{cyburt04}. 

We summarize our procedure for analyzing single data sets as follows.
\begin{enumerate}
\item
We find best fits and statistical uncertainties, where the statistical
errors of the data points dominate the $\chi^2$ analysis.
\item
The total normalization error is the quadratic sum of the intrinsic
normalization error and our quality-of-fit measure, the discrepancy
error.
\end{enumerate}

We are then left with the remaining question of how to treat multiple
data sets. We discuss two methods. The first method, adopted
in~\cite{cyburt04} treats the data sets as totally correlated and
comprising a single data set. The second method treats the data sets
as if they were completely independent. Reality is likely between
these two possibilities; however at present we have no good
prescription for determining how correlated data sets actually are, and
therefore present results for these two limiting cases. Additionally, both
methods yield similar results, suggesting that the methods are accurate
and robust. Generally, the totally correlated method yields more
conservative uncertainties as the intrinsic normalizations are not
treated statistically.

\subsection{Completely Correlated Data Sets}

Assuming that individual data sets are totally correlated simplifies
the analysis. Since the exact nature of the correlation is unknown, we
cannot rigorously define a correlation matrix, so we continue
to use the definition in Eq.~\ref{eqn:datcov} ($C_{i_n,j_m} =
\delta_{n,m}C_{i_n,j_n}$) and its inverse in Eq.~\ref{eqn:invdatcov}
($C_{i_n,j_m}^{-1} = \delta_{n,m}C_{i_n,j_n}^{-1}$), generalized for
multiple data sets. By virtue of the large data set limit for the
inverse covariance matrix, we are led to the conclusion that the
precise nature of the correlations is relatively unimportant, as the statistical
uncertainties dominate the $\chi^2$ analysis.

As before, the total normalization error is the quadratic
sum of the intrinsic normalization error and the discrepancy error
($\epsilon_{tot}^2=\epsilon_{norm}^2+\epsilon_{disc}^2$). The
discrepancy error as defined in Eq.~\ref{eqn:disc} is valid for a
single data set and hence also for the case of totally correlated data sets. Since the individual
data sets are completely correlated, the overall normalization error must be
some average of the individual normalization errors. We adopt the
normalization error prescription of Ref.\ \cite{cyburt04}:
\begin{equation}
\label{eqn:norm}
\epsilon_{norm}^2 = \frac{\sum_n \frac{\epsilon_n^2}{\chi^2_n}}{\sum_n \frac{1}{\chi^2_n}},
\end{equation}
where $\epsilon_n$ are the individual data set normalization errors,
and $\chi^2_n$ is the $\chi^2$ per datum of data set $n$ with respect to the best
fit. This weighting scheme gives more weight to data sets that
agree with the best fit model. We also point out that this
normalization error assignment is bounded by the smallest and largest
normalization errors and is not reduced by the number of data sets.
This reflects the fact that the data sets are completely correlated
and the normalization errors cannot be treated statistically in this
case.

The expectation value and statistical variance of the best fit model
are denoted $E[y(x)]$ and $V_S[y(x)]$, respectively. Including the
normalization error, the total variance in the best fit model is
$V_T[y(x)] = (1+\epsilon_{tot}^2)V_S[y(x)] +
\epsilon_{tot}^2E[y(x)]^2$. This completes our description of the formalism for completely correlated data sets, which we refer to as the correlated normalization analysis.

\subsection{Completely Independent Data Sets}

If data sets are truly independent from each other, we can treat the
normalizations statistically, once best fits and statistical errors
are found for each data set. We consider only linear models, for
which $y(x) = \sum_{p=1}^P a_p {\rm Y}_p(x)$, where the ${\rm Y}_p(x)$ are
known functions, and follow the prescription laid out in the previous
section for each data set. The expectation value and statistical
variance of $y(x)$ are $E[y(x)] = \sum_{p=1}^P \hat{a}_p {\rm Y}_p(x)$
and $V_S[y(x)] = \sum_{p,q=1}^P {\mathcal C}_{p,q} {\rm Y}_p(x){\rm
Y}_q(x)$, respectively. With the best fit parameters
$\hat{a}_p^{(n)}$ and parameter covariance ${\mathcal C}_{p,q}^{(n)}$
for each data set $n$ in hand, we can combine individual data sets and
find a global best fit. Upon minimization, the $\chi^2$ can be
decomposed as:
\begin{equation}
\label{eqn:parchi2}
\chi^2 = \chi^2_{min} + \sum_n \sum_{p,q=1}^P {{\mathcal C}_{p,q}^{(n)}}^{-1} (a_p - \hat{a}_p^{(n)}) (a_q - \hat{a}_q^{(n)}),
\end{equation}
where the calligraphic ${\mathcal C^{(n)}_{p,q}}$ is the covariance
between the $p^{th}$ and $q^{th}$ parameters of the $n^{th}$ data set, with best fit parameters
given by $\hat{a}_p^{(n)}$, provided there are at least as many data points
$N$ as fitting parameters $P$. Note that with linear models all
parameters are gaussian, i.e., the $\chi^2$ is a quadratic function of the
parameters. 

In order to combine multiple data sets, we must first
replace the statistical variance, $V_S[y(x)]$, with the total variance
$V_T[y(x)]$. In other words, we replace the parameter covariance matrix
${\mathcal C}_{p,q}^{(n)}$ for each data set $n$ with
$(1+\epsilon_{tot,n}^2){\mathcal C}_{p,q}^{(n)} +
\epsilon_{tot,n}^2\hat{a}_p^{(n)}\hat{a}_q^{(n)}$. Inserting this 
new parameter covariance into Eq.~\ref{eqn:parchi2}, we minimize the
$\chi^2$ to find the best fit parameters and their statistical
covariances, which now include the normalization errors of each
data set.

We also need to quantify how well data sets agree with each other. The
variance of the best global fit contains no information about the mutual consistency
 of data sets. This is seen if we shift the mean
values of data points in a single data set, but keep the same absolute
uncertainties. The parameter covariance is not changed, even though
data sets can be severely discrepant. Thus we need to define a global
discrepancy error. To do this, we calculate the renormalizations
needed for the global best fit to minimize the $\chi^2$ for individual
data sets. The dispersion of these renormalizations provides
an estimate of the discrepancy between data sets. We find
renormalizations $\alpha_n$ and their total errors $\sigma_n$ for
each data set and then their dispersion according to
\begin{equation}
\epsilon_{disc}^2 = \frac{\sum_n \left(\frac{\alpha_n-1}{\sigma_n}\right)^2}{\sum_n \frac{1}{\sigma_n^2}}.
\end{equation}
This definition is similar to that discussed for the individual
data sets, except that here no correlations exist between data sets.
Given the global best fit $E[y(x)]$ and its statistical variance
$V_S[y(x)]$, the total variance in this case is given by: $V_T[y(x)] =
(1+\epsilon_{disc}^2)V_S[y(x)] + \epsilon_{disc}^2E[y(x)]^2$. We refer to this formalism as the independent normalization analysis.

\section{The Data}

With these formalisms in place, we now discuss the data available to
constrain the astrophysical $S$ factor of the $^7$Be(p,$\gamma$)$^8$B
reaction. In some cases there is sufficient reason to exclude data sets
from the analysis. We consider only low energy data,
$E_{cm}<425$ keV, when determining the best fits to $S_{17}(E)$ as nuclear structure uncertainties complicate and render more uncertain the extrapolation when higher energy data are included \cite{jennings98,davids03,junghans03}.

\subsection{Radiative Capture Data}

Initially, we considered the data sets of Kavanagh~\cite{kavanagh60},
Parker~\cite{parker66}, Vaughn \emph{et al.}~\cite{vaughn70},
Filippone \emph{et al.}~\cite{filippone83}, Strieder \emph{et al.}~\cite{strieder01}, Hammache-1~\cite{hammache98}, Hammache-2~\cite{hammache01},
Hass \emph{et al.}~\cite{hass99}, Junghans-BE1~\cite{junghans02},
Junghans-BE3~\cite{junghans03}, and Baby \emph{et al.}~\cite{baby03a}. We exclude the data of Kavanagh~\cite{kavanagh60}, Parker~\cite{parker66} and
Vaughn \emph{et al.}~\cite{vaughn70} because these authors do not present enough information
to adequately determine a normalization error. We do not use the measurement of
Hass \emph{et al.}~\cite{hass99} simply because the data lie above our 425 keV
energy cutoff.

Several details of our analysis bear mention here. One of the two target thickness determinations in the Filippone \emph{et al.} measurement \cite{filippone83} relies on the $^7$Li(d,p) reaction. We adopt the recommendation of \cite{adelberger98} for the value of this reaction cross section.
The Hammache-2~\cite{hammache01} data consist of 3 points, two of
which are measured relative to the third. Ideally, one would like to
include all 3 points, but not enough information is given on the
third point to determine an intrinsic normalization error. We thus
adopt the 2 relative measurements as the data set, using the third to
determine the normalization error. Ref.\ \cite{junghans03} presents data from their BE1 measurement renormalized using the BE3 data. These renormalized BE1 data are not independent of the BE3 data. Therefore,
we consider here the BE3 data \cite{junghans03} and the original BE1 data~\cite{junghans02}, which are independent. Finally we note that there is some discussion in the literature \cite{junghans03} regarding the uncertainties in the data of Baby \emph{et al.} \cite{baby03a}. We take the uncertainties for this measurement from Table II of Ref.\ \cite{baby03a}.

\subsection{Coulomb Dissociation Data}

We considered the Coulomb dissociation (CD) data of
Kikuchi \emph{et al.}~\cite{kikuchi97,kikuchi98}, Iwasa \emph{et al.}~\cite{iwasa99},
Sch\"umann \emph{et al.}~\cite{schuemann03} and Davids \emph{et al.}~\cite{davids01,davids03}. We exclude the measurements of Kikuchi \emph{et al.}~\cite{kikuchi97,kikuchi98} and Iwasa \emph{et al.}~\cite{iwasa99} due to concerns over the way these data were analyzed. The data from these measurements were not analyzed using a cut on the maximum scattering angle of the $^8$B center-of-mass, corresponding semiclassically to a minimum impact parameter. This means that the effects of nuclear absorption, diffraction dissociation, and $E2$ transitions are present in the data, and that the inferred $E1~S$ factors may not be reliable. Since we lack the detailed experimental information required to correct for these effects, we do not consider these data here. Hence we include only the CD data of Sch\"umann \emph{et al.}~\cite{schuemann03} and Davids \emph{et al.}~\cite{davids01,davids03}, which were analyzed using scattering angle cuts to minimize the nuclear and $E2$ contributions and their associated uncertainties.

\section{Theoretical Structure Models}\label{models}

In general, two classes of models have been used to describe the
structure of $^8$B, single particle potential models that treat $^8$B
as a $p$-wave proton coupled to a $^7$Be core in its ground state,
and microscopic cluster models that include two configurations of the
three clusters $^3$He, $\alpha$, and a proton. Recently, most
experimenters have used the cluster model of Descouvemont and Baye
(DB) \cite{db94} to extrapolate their data to zero energy. This
generator coordinate method employs a central nucleon-nucleon
interaction along with Coulomb and spin-orbit
interactions. Ref.\ \cite{db94} used the Volkov II nucleon-nucleon
interaction \cite{volkov65}, while a more recent preliminary effort by
Descouvemont and Dufour (DD) \cite{dd04} opts for the Minnesota force
\cite{mn}, which describes low mass systems better. These cluster models include
excited $^7$Be configurations. The single particle potential models
generally employ a central Woods-Saxon + Coulomb potential, and
optionally a spin-orbit interaction, which can be neglected in
calculations of the $E1$ $S$ factor provided the central potential depth
is properly adjusted. In these models, the $p$-wave potential depth is
fixed by the $^8$B binding energy. The depths for the other partial
waves can be chosen identically, but a better choice is to fix them
using the well-measured $s$-wave scattering lengths for channel spin 1
and 2 in the isospin mirror system $^7$Li + $n$ \cite{koester83}. The
scattering lengths in the $^7$Be + $p$ system have also been measured
\cite{angulo03}, but with much lower precision. The scattering lengths for the dominant $S$ = 2 channel are consistent between the isospin mirrors, but there is a 2.7$\sigma$ discrepancy in the value for the $S$ = 1 channel. This discrepancy is not understood at present, and deserves attention in the future. In this work, we
consider the DD cluster model and the $^7$Li + $n$ potential model of Davids and
Typel (DT) \cite{davids03}, which reproduces the $^7$Li + $n$ scattering
lengths. These models well represent their respective classes.

\begin{figure}
\includegraphics[width=\linewidth]{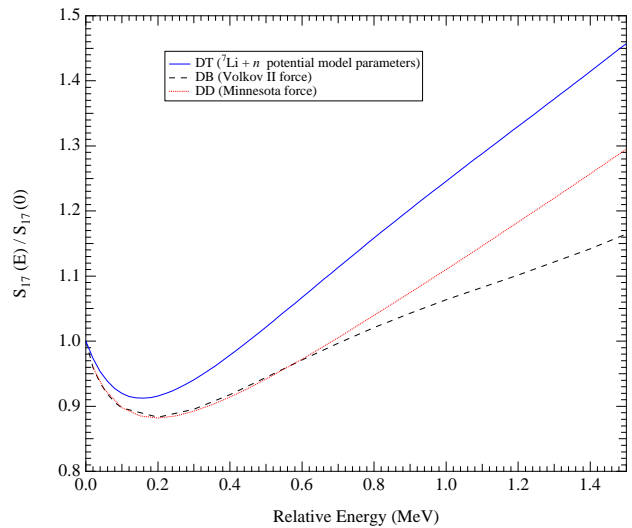}
\caption{\label{fig1}(Color online) Shapes of the $S$ factors predicted by the Descouvemont and Baye (DB) and Descouvemont and Dufour (DD) cluster models and 
the Davids and Typel (DT) potential model as a function of relative energy (E$_{cm}$). Shown is the ratio of the $E1$ $S$ factor to its
value at E$_{cm}$ = 0, which allows the shapes of models with different
absolute normalizations to be shown in the same figure. The shapes of
the two cluster models are very similar at low energies, but substantial
deviations exist above 700 keV. The potential model is seen to have a
slightly larger slope at high energies, but also a very different
shape below 400 keV.}
\end{figure}

Both of these models have virtues. The cluster model allows more
configurations than does the potential model, and therefore might be
expected to describe the physics better. On the other hand, the
potential model is simple, and has been tuned to reproduce the elastic
scattering data. Fig.\ \ref{fig1} shows the shapes predicted by the
potential and cluster models.

\section{Structure-Model-Dependent Analysis}

With the theoretical structure models described briefly in Section \ref{models}, our
fitting procedure involves only a single parameter, an absolute
normalization. We fit data using the DT $^7$Li + $n$ potential model and the DD cluster model employing the
formalism presented in Section \ref{fit}. Our results are summarized in
Tables~\ref{tab:fitli7425} and \ref{tab:fitDBM425}. 

\begin{table}
\begin{center}
\caption{Best fits to radiative capture (RC) and Coulomb dissociation (CD) data for 
$E_{cm}<425$ keV using the DT $^7$Li + $n$ potential model.
Shown are the best fit astrophysical $S$ factors and their standard
deviations (statistical) at $E_{cm}=0$. Also shown are the two
individual contributions to the total systematic error, the
intrinsic normalization error and the discrepancy error, given in
percent. An additional potential model parameter uncertainty should be
added to the other systematic errors, with $\epsilon_{model}$ = 0.01\%, reflecting
the small uncertainties in the $^7Li+n$ elastic scattering data.\label{tab:fitli7425}}
\begin{tabular}{||c|l|c|c|c||}\hline\hline
  & Data set (\# of points)& $S_{17}(0)$ (eV b)& $\epsilon_{norm}$ & $\epsilon_{disc}$ \\ \hline 
  & Filippone  (6)& $19.3\pm0.5$ & 11.9\% & 5.6\% \\ 
  & Strieder (2)& $17.9\pm0.6$ & 8.3\% & 2.7\% \\ 
  & Hammache-1 (3)& $19.4\pm0.6$ & 4.9\% & 5.7\% \\ 
  Direct & Hammache-2 (2)& $18.8\pm1.2$ & 12.2\% & 8.1\% \\ 
  & Junghans-BE1 (8)& $21.6\pm0.2$ & 2.7\% & 0.7\% \\
  & Junghans-BE3 (13)& $21.2\pm0.1$ & 2.3\% & 1.3\% \\ 
  & Baby (3)& $19.8\pm0.7$ & 2.2\% & 2.2\% \\
\hline
  & Davids (2)& $16.6\pm0.5$ & 7.1\% & 0.1\% \\
 CD  & Sch\"umann (2)& $18.4\pm0.8$ & 5.6\% & 0.8\% \\ 
\hline\hline
  & All RC but Junghans & $19.3\pm0.7$ & --- & 3.8\%  \\
 indep. & Junghans& $21.4\pm0.4$ & --- & 0.8\%  \\
 norm.& All radiative capture & $20.8\pm0.4$ & --- & 5.1\%  \\
  & Coulomb dissociation & $17.5\pm0.9$ & --- & 5.5\%  \\
\hline
  & All but Junghans & $19.1\pm0.3$ & 8.5\% & 6.2\% \\
 corr. & Junghans & $21.3\pm0.1$ & 2.6\% & 1.4\% \\
 norm. & All radiative capture & $21.2\pm0.1$ & 5.6\% & 3.2\% \\
  & Coulomb dissociation & $17.1\pm0.5$ & 6.7\% & 5.5\% \\
\hline\hline
\end{tabular}
\end{center}
\end{table}

\begin{table}
\begin{center}
\caption{Same as table~\ref{tab:fitli7425}, except for the DD cluster
model. A model uncertainty should be added to the other systematic
errors, but DD do not present formal errors for this model.
\label{tab:fitDBM425}}
\begin{tabular}{||c|l|c|c|c||}\hline\hline
 & Data set & $S_{17}(0)$ (eV b)& $\epsilon_{norm}$ & $\epsilon_{disc}$ \\ \hline 
  & Filippone  & $20.1\pm0.5$ & 11.9\% & 6.1\% \\
  & Strieder & $18.8\pm0.7$ & 8.3\% & 2.4\% \\
  & Hammache-1  & $20.4\pm0.7$ & 4.9\% & 5.9\% \\
Direct & Hammache-2  & $19.1\pm1.2$ & 12.2\% & 8.2\% \\
  & Junghans-BE1 & $22.6\pm0.2$ & 2.7\% & 1.0\% \\
  & Junghans-BE3 & $22.1\pm0.1$ & 2.3\% & 1.4\% \\
  & Baby   & $20.9\pm0.7$ & 2.2\% & 2.7\% \\ 
\hline
  & Davids  & $17.4\pm0.6$ & 7.1\% & 1.0\% \\
 CD  & Sch\"umann  & $19.2\pm0.9$ & 5.6\% & 1.6\% \\ 
\hline\hline
  & All RC but Junghans & $20.3\pm0.7$ & --- & 4.2\% \\
 indep. & Junghans & $22.3\pm0.5$ & --- & 1.1\% \\
 norm.& All radiative capture & $21.8\pm0.4$ & --- & 4.9\% \\
  & Coulomb dissociation & $18.2\pm1.0$ & --- & 5.3\% \\
\hline
  & All RC but Junghans & $20.0\pm0.3$ & 8.6\% & 6.5\% \\
corr. & Junghans& $22.2\pm0.1$ & 2.5\% & 1.6\% \\
 norm. & All radiative capture& $22.1\pm0.1$ & 5.5\% & 3.3\% \\
  & Coulomb dissociation & $17.9\pm0.5$ & 6.7\% & 5.5\% \\
\hline\hline
\end{tabular}
\end{center}
\end{table}

We first derive best fits for individual experiments. The $S_{17}(0)$
determinations are in excellent agreement with previous
analyses~\cite{junghans03,davids03}. In general, the discrepancy errors
are smaller than or comparable to the intrinsic normalization errors
of each data set. Note that the central values of $S_{17}(0)$ for
the radiative capture data range from 18-20 eV b and 21-22 eV b for non-Junghans and Junghans data sets respectively, while the CD
data lie in the range of 16-19 eV b. This hints at some level of
disagreement among the radiative capture data sets, especially when comparing the two
Junghans experiments~\cite{junghans02,junghans03} with the other
radiative capture data, in addition to that between the radiative capture and the CD
data.

To further explore and quantify this disagreement we look at different
combinations of the data in our multiple experiment fits.
We first look at the radiative capture data alone. In these model-dependent
analyses, we consider 3 combinations of the radiative capture data, (1) all but the Junghans data, (2) only Junghans data, and (3) all the radiative capture data. Using the DT $^7$Li + $n$ potential model, the independent normalization method gives for $S_{17}(0)$ in eV b: (1) $19.3\pm1.0$, (2) $21.4\pm0.5$ and (3)
$20.8\pm1.1$; using the correlated normalization method, we
find (1) $19.1\pm2.0$, (2) $21.3\pm0.7$ and (3) $21.2\pm1.4$ eV b. For the CD data we find
$17.5\pm1.3$ and $17.1\pm1.6$ eV b using the independent normalization and correlated normalization methods respectively.

We find that the CD data and the non-Junghans radiative capture data are
consistent, showing approximately a 1$\sigma$ discrepancy, while the CD data
and Junghans data are discrepant at a little over the 2$\sigma$ level. Furthermore there
is a 1 to 2$\sigma$ disagreement between the Junghans data and the
other radiative capture measurements. The Junghans data dominate
the fit when combining all the radiative capture data, due to their extremely
small errors and the size of the data sets. Similar but somewhat higher results are obtained with the DD cluster model.
Interestingly, the high precision Junghans data are described slightly better by the DT
$^7$Li + $n$ potential model than by the DD cluster model, as seen in our
quality-of-fit measure, the discrepancy error, but not at a
significant level. The nature of the discrepancies between the
Junghans data and both the non-Junghans radiative capture data and the CD data
must be understood before we can address which model describes the data
best. In an attempt to go beyond structure-model-dependent results, we now explore a structure-model-independent analysis.

\section{Pole Model Analysis}

Performing a structure-model-independent analysis
will provide insight into the
quality of existing data for the $^7$Be(p,$\gamma$)$^8$B reaction
below 425 keV. We use the expansion suggested
by~\cite{jennings98}, adopting a three-parameter model:
\begin{equation}
\label{eqn:mif}
S_{17}(E) = S_{17}(0) + \alpha\frac{E}{Q(E+Q)} + \beta E,
\end{equation}
where $Q=137.5$ keV. In addition to terms constant and linear in energy, this functional form contains a simple pole term, a universal feature of radiative capture reactions that is independent of the details of nuclear structure models. We have chosen a fit that is linear in its
parameters, but one can easily calculate the non-linear parameters adopted in
\cite{jennings98}: the pole term is given by 
$a=-\alpha/S_{17}(0)$ and the slope term by $c=\beta/S_{17}(0)$. A data set must have at least 3 data points in
order for us to employ this three-parameter fit. Our results are summarized in
Table~\ref{tab:mifit}. Of all the data, only the
most recent Junghans-BE3 data set~\cite{junghans03} provides a significant structure-model-independent
constraint on the low energy behavior of the astrophysical $S$ factor,
implying $S_{17}(0)=24.3\pm4.0$ eV b. With this data set providing
the only significant low-energy constraint on $S_{17}$, it dominates
the fit when we combine the radiative capture data to find $23.6\pm3.4$
eV~b for the Junghans data and $23.4\pm3.4$ eV~b for all the radiative capture data,
using the independent normalization method. Using the
correlated normalization method we find $24.1\pm3.4$ eV~b for
Junghans alone, and $18.9\pm3.2$ eV~b if we include all
data sets. With the current data we
can place robust constraints on the extrapolated value
$S_{17}(0)$ independent of any particular structure model. We find that the central value is calculated to
lie between 19-24 eV~b with a total uncertainty of $\pm3.5$ eV~b.

\begin{table*}[h]
\begin{center}
\caption{Best fit parameters for data with $E_{cm}<425$ 
keV with the pole model functional form of Eq.~\ref{eqn:mif}.
\label{tab:mifit}}
\begin{tabular}{||c|l|c|c|c||c|c||}\hline\hline
  &  Data & $S_{17}(0)$ (eV b) & $\alpha$ (eV b MeV)  & $\beta$ (eV b MeV$^{-1}$) & $\epsilon_{norm}$ & $\epsilon_{disc}$ \\ \hline
  & Fillipone  & $38.6\pm15.7$ & $-8.5\pm5.4$ & $70.6\pm37.2$ & 11.9\% & 3.1\% \\
  & Hammache-1  & $-2430\pm2160$  & $-623\pm554$ & $-2308\pm2103$  & 4.9\% & 0.0\% \\
RC & Junghans-BE1 & $18.4\pm10.6$ & $-0.0\pm3.2$ & $7.1\pm17.5$ & 2.7\% & 0.7\% \\
  & Junghans-BE3 & $24.3\pm3.9$  & $-2.0\pm1.3$  & $18\pm9$   & 2.3\% & 1.3\% \\
  & Baby   & $55.3\pm213.1$  & $-11.4\pm56.2$  & $66\pm229$ & 2.2\% & 0.0\% \\ 
\hline\hline
  & All RC but Junghans & $34.9\pm13.2$ & $-7.2\pm4.3$  & $60.2\pm27.4$ & ---  & 1.3\% \\
  indep. & Junghans& $23.6\pm3.4$  & $-1.7\pm1.1$  & $16.1\pm7.1$  & ---  & 0.7\% \\
norm.& All radiative capture & $23.4\pm3.3$  & $-1.8\pm1.1$  & $16.6\pm6.8$  & ---  & 3.7\% \\
  & Coulomb dissociation & --- & --- & --- & --- & --- \\
\hline
  & All RC but Junghans \footnote{All non-Junghans RC data sets with at least 3 points}& $35.4\pm13.0$ & $-7.1\pm4.2$ & $57.2\pm26.8$  & 6.7\% & 3.6\% \\
  & All but Junghans \footnote{All non-Junghans RC}& $29.0\pm7.9$ & $-5.0\pm2.9$ & $43.5\pm21.2$  & 7.7\% & 5.4\% \\
  corr. & Junghans& $24.1\pm3.4$  & $-1.9\pm1.1$ & $18.5\pm6.9$  & 2.6\% & 1.3\% \\
  norm. & All radiative capture  & $18.9\pm2.9$  & $-0.1\pm1.0$ & $5.7\pm6.1$ & 6.0\% & 3.2\% \\
  & Coulomb dissociation & $-13.5\pm26.4$ & $9.3\pm8.6$  & $-53\pm53$ & 6.5\% & 3.6\% \\
\hline\hline
\end{tabular}
\end{center}
\end{table*}

We see that in order to place a tighter constraint on $S_{17}$, one
must assume some low-energy behavior. We now explore a
constrained pole model which exploits knowledge of the
astrophysical $S$ factor around its subthreshold pole, on which all
theories agree.

\section{Constrained Pole Model Analysis}

In order to pin down the $S$ factor with higher precision, we need to
examine the available theories and determine what, if any, universal
information is available. To do this we fit each theory, normalized
such that $S_{17}(0)\equiv 1$, using the pole model form, which now reduces to
a 2 parameter fit, depending on the pole term $a=-\alpha/S_{17}(0)$
and the slope term $c=\beta/S_{17}(0)$. We fit to 4 theories, the two
potential models of DT and the cluster-model calculations of DB and DD. Our results are
summarized in Table~\ref{tab:thryfit}. 

\begin{table}[htb]
\begin{center}
\caption{Best fit parameters for several theoretical structure models below $E_{cm}=425$ 
keV.
\label{tab:thryfit}}
\begin{tabular}{||l|c|c||}\hline\hline
Model   & $a$ (keV) & $c$ (MeV$^{-1}$) \\ \hline
DT $^7$Li + $n$ potential & $45.7\pm0.5$ & $0.553\pm0.007$ \\ \hline
DT $^7$Be + $p$ potential & $45.0\pm0.6$ & $0.433\pm0.007$ \\ \hline
DB Volkov II& $45.5\pm0.1$ & $0.434\pm0.001$ \\ \hline
DD Minnesota  & $44.5\pm0.1$ & $0.404\pm0.002$ \\ \hline\hline
\end{tabular}
\end{center}
\end{table}

As one can see the pole term
is robustly determined, with $a \in[44,46]$ keV. We adopt
$a=45$ keV as our canonical value, and allow this information
to propagate through our data analysis to see how our
constraints on the low-energy behavior of $S_{17}$ improve. We thus  
use the two-parameter fit:
\begin{equation}
\label{eqn:qmif}
S_{17}(E) = S_{17}(0)\left[1 - a\frac{E}{Q(E+Q)}\right] + \beta E,
\end{equation}
with the parameter $a$ being fixed at 45 keV. Our results are
summarized in Table~\ref{tab:fit425a45}.

\begin{table*}[htb]
\begin{center}
\caption{Results of a constrained pole model fit of radiative capture and Coulomb dissociation data below $E_{cm}=425$ 
keV. Shown are the
best fit astrophysical $S$ factors, the associated slope parameter
$\beta$, and their standard deviations (statistical). Also shown are
both the intrinsic normalization errors and the discrepancy
normalization errors for combined data sets, all cited in percent,
which should be added in quadrature with the statistical errors to get
the total error. An additional systematic error must be added
due to the uncertainty in $a$; $\Delta S_{17}(0)/S_{17}(0) \approx
0.16\Delta a/a$, and $\Delta\beta/\beta \approx 0.80\Delta a/a$, which for
$a=45\pm1$ keV yields $\Delta S_{17}(0)/S_{17}(0) = 0.4\%$ and
$\Delta\beta/\beta = 1.8\%$, respectively.
\label{tab:fit425a45}}
\begin{tabular}{||c|l|c|c|c|c|c|c|c|c||}\hline\hline
  & Data set & $S_{17}(0)$ (eV b)& $\beta$ (eV b MeV$^{-1}$) & $\epsilon_{norm}$ &
$\epsilon_{disc}$ \\ \hline 
  & Filippone  & $16.4\pm2.8$ & $18.1\pm7.2$  & 11.9\% & 4.9\% \\ 
  & Strieder   & $30.6\pm17.2$ & $-17.8\pm37.6$ & 8.3\% & 4.0\% \\ 
& Hammache-1  & $-2.7\pm16.7$ & $56.2\pm34.4$ & 4.9\% & 3.7\% \\ 
RC  & Hammache-2  & $-2.3\pm16.1$ & $159.\pm113.$ & 12.2\% & 0.0\% \\ 
  & Junghans-BE1 & $21.4\pm1.1$ & $12.1\pm2.6$  & 2.7\% & 0.7\% \\
  & Junghans-BE3 & $21.4\pm0.6$ & $11.3\pm1.6$  & 2.3\% & 1.3\% \\ 
  & Baby   & $14.6\pm8.5$ & $21.9\pm17.8$ & 2.2\% & 0.7\% \\
\hline
  & Davids  & $16.6\pm2.5$ & $9.1\pm6.4$  & 7.1\% & 0.0\% \\
 CD  & Sch\"umann  & $17.5\pm4.9$ & $12.5\pm13.2$ & 5.6\% & 0.0\% \\ 
\hline\hline
  & All RC but Junghans & $16.3\pm2.3$ & $17.3\pm5.0$ & --- & 3.5\% \\
  indep. & Junghans  & $21.4\pm0.7$ & $11.6\pm1.4$ & --- & 0.8\% \\
norm.& All radiative capture & $20.7\pm0.6$ & $11.3\pm1.3$ & --- & 4.8\% \\
  & Coulomb dissociation & $17.5\pm2.3$ & $9.7\pm5.8$  & --- & 5.1\% \\
\hline
  & All RC but Junghans & $17.9\pm1.4$ & $13.4\pm3.3$ & 8.0\% & 5.8\% \\
corr. & Junghans  & $21.1\pm0.5$ & $12.3\pm1.3$ & 2.6\% & 1.3\% \\
 norm. & All radiative capture & $21.4\pm0.5$ & $11.0\pm1.2$ & 5.7\% & 3.2\% \\
  & Coulomb dissociation  & $17.4\pm2.2$ & $8.7\pm5.7$  & 6.4\% & 4.8\% \\
\hline\hline
\end{tabular}
\end{center}
\end{table*}

Using this information, individual experiments determine
$S_{17}(0)$ much more precisely than in the unconstrained pole model fit.
Again the Junghans data provide the strongest constraints,
$21.4\pm1.3$ eV~b~\cite{junghans02} and $21.4\pm0.8$
eV~b~\cite{junghans03} respectively. However, the discrepancy between
the two Junghans data sets and the other radiative capture experiments is
quite apparent, with the fit to the combined Junghans data giving $21.4\pm0.7$ eV~b and
the other radiative capture data yielding $16.3\pm2.4$ eV~b. This is a 2$\sigma$
discrepancy between these two low-energy extrapolations, using the
independent normalization method. This tension is somewhat reduced in
the correlated normalization method, which yields
$21.1\pm0.8$ eV~b and $17.9\pm2.3$ eV~b for the Junghans and
non-Junghans radiative capture data sets respectively, a discrepancy slightly more than 1$\sigma$. The CD data yield
$17.5\pm2.5$ eV~b and $17.4\pm2.6$ eV~b using the independent normalization
and correlated normalization methods, respectively.
It is quite remarkable that the non-Junghans radiative capture data and the CD data agree so
well, while both disagree with the Junghans data.
This may be due to the sparseness of the data in these
data sets, but the deviations from the Junghans results are
significant. It would be desirable to have an independent confirmation of the Junghans measurements, since they deviate significantly from the other data and dominate the central value of combined fits.

At this time, it is unclear what is causing the discrepancies.
Arguably, the Coulomb dissociation measurements have the least in
common with the radiative capture measurements and large systematics of their
own, so one could claim that they are the source of the discrepancy.
However, this fails to explain the discrepancy among radiative capture
measurements, namely between the two Junghans
experiments~\cite{junghans02,junghans03} and that of
Filippone~\cite{filippone83}, which dominates the non-Junghans
data sets. As shown in Table~\ref{tab:fit425a45}, none of the other
radiative capture measurements below $E_{cm}=425$ keV provides a significant
constraint on $S_{17}(0)$. It is unclear which experiments are
responsible for the discrepancy. 

With these remaining uncertainties in mind, it is helpful to remind
ourselves that we have defined a rigorous treatment for exactly these
kinds of discrepancies. Our treatment has examined the
level of concordance and quantified it in terms of a discrepancy
error. We
thus recommend an astrophysical $S$ factor at zero energy of:
\begin{eqnarray}\label{indepnormeq}
S_{17}(0) = 20.7\pm1.2 {\rm\ eV~b} \mbox{\ \ \rm indep. normalization} \\
S_{17}(0) = 21.4\pm1.4 {\rm\ eV~b} \mbox{\ \ \rm corr. normalization}
\end{eqnarray}
for the radiative capture data, and 
\begin{eqnarray}
S_{17}(0) = 17.5\pm2.5 {\rm\ eV~b} \mbox{\ \ \rm indep. normalization} \\
S_{17}(0) = 17.4\pm2.6 {\rm\ eV~b} \mbox{\ \ \rm corr. normalization}
\end{eqnarray}
for the CD data.

One can compare our low-energy $S$ factor determinations with those
determined from asymptotic normalization coefficients (ANCs). There
are measurements from (1) proton
transfer reactions~\cite{azhari01}, (2) $^8$B breakup
reactions~\cite{trache04} and (3) neutron transfer
reactions~\cite{trache03}. The asymptotic normalization coefficient can be very simply related to the astrophysical $S$ factor, so we quote the measurements in these terms. These determinations yield (1)
$17.3\pm1.8$~eV b, (2) $18.7\pm1.9$~eV b and (3) $17.6\pm1.7$~eV b for
the astrophysical $S$ factor of the $^7$Be(p,$\gamma$)$^8$B reaction.
These values of $S_{17}(0)$ agree perfectly with the CD data and
non-Junghans radiative capture data. The ANC-derived $S$ factors disagree
with the Junghans data at slightly more than the 1$\sigma$ level.
Again, these discrepancies need to be more fully explored to better determine
the low-energy behavior of the $S$ factor.

Thus far we have primarily discussed the low-energy extrapolations of
the $S$ factor $S_{17}(0)$ and not the slope. Using our
constrained pole model we find that the slopes based on
the non-Junghans radiative capture, Junghans, and CD data are all roughly consistent
with each other. In fact, the Junghans and CD slopes agree
remarkable well, though the CD data have sizable errors. The
non-Junghans radiative capture slope disagrees at the 1$\sigma$ level with both the
Junghans and the CD data. Again, the non-Junghans fit is dominated
by the Filippone~\cite{filippone83} data, with substantial
uncertainties. These discrepancies disappear when one uses the
correlated normalization method, suggesting that no significant
deviation in the slope is observed and that the constraint on the
slope using $E_{cm}<425$ keV data is not particularly strong. The
fact that the statistical errors in this parameter dominate over the
normalization error supports this.

\section{\label{concl}Conclusions}

We have presented a robust formalism for fitting data that both
properly propagates known systematic uncertainties and quantifies
the quality of fit, incorporating a discrepancy error into the total
systematic error. We discuss two limiting cases, one in which data sets are considered completely independent from each other, and a second in which all data sets are totally correlated.
These two methods, the independent normalization and correlated
normalization methods, yield similar results, providing robust constraints
and suggesting that most previous analyses have underestimated the true
uncertainty. 

A structure-model-dependent analysis was performed using the DT
$^7$Li + $n$ potential~\cite{davids03} and the DD Minnesota force
cluster~\cite{dd04} models. The $^7$Li + $n$ potential model generally
predicts low-energy extrapolated $S$ factors smaller than the
Minnesota-interaction cluster model. With the available data, no
significant preference for one model is observed. We explored
a structure-model-independent fit to the data, finding that only the
Junghans~\cite{junghans03} data placed even modest constraints on the
low-energy $S$ factor. Identifying a feature common to all
models, the relative strength of the subthreshold pole term, we reduced the extrapolation error considerably. Even though we
find evidence for discrepancies between the Junghans \emph{et al.} radiative capture measurements and the others~\cite{filippone83,strieder01,hammache01,hammache98,baby03a}),
our rigorous and careful treatment of systematic errors provides a
robust determination of $S_{17}(0)$. Our analysis of indirect Coulomb dissociation data \cite{schuemann03,davids03} and ANC determinations \cite{azhari01,trache03,trache04} found mutual agreement and consistency with the radiative capture measurements other than those of Junghans \emph{et al.}

The dominant source of error in the standard solar model predictions
for the total $^8$B neutrino flux is the uncertainty in the heavy
metal abundance. The dominant nuclear uncertainties stem from
uncertainties in the $^3$He($\alpha$,$\gamma$)$^7$Be and
$^7$Be(p,$\gamma$)$^8$B reactions. Using a technique similar to that employed here, the authors of \cite{cyburt04} find a total error in the $S_{34}$ normalization
of 17\%. If we adopt our determination of $S_{17}(0)$ using the independent normalization method (Eq.\ \ref{indepnormeq}) and the $S_{34}$ error assignment from \cite{cyburt04} we find the following
standard solar model (BP04)~\cite{bahcall04} prediction for the total
$^8$B solar neutrino flux in units of $10^{6}$ cm$^{-2}$ s$^{-1}$:
\begin{equation}
\phi(^8{\rm B}) = 5.63\left[1\pm0.058(S_{17})\pm0.15(S_{34})\pm0.21\right].
\end{equation} Here we have separated the individual contributions to
the total error in the neutrino flux, those from $S_{17}$, $S_{34}$,
and the other standard solar model parameters, which when added in
quadrature yield a total error of 26\%. We can see
that the new $S_{17}$ error assignment contributes 6\% to the total
neutrino flux error.

While S$_{17}$ now makes only a relatively small contribution to the total uncertainty in the predicted $^8$B solar neutrino flux, ongoing measurements of S$_{34}$ and improved radiative opacity tables may reduce the other solar model uncertainties substantially in the near future. In order to further reduce the uncertainty on S$_{17}$, a new high precision measurement with a detailed error budget would be required.

\section{Acknowledgements}

This work was supported by the Natural Sciences and Engineering
Research Council of Canada. We acknowledges helpful discussions with Shung-ichi Ando, Sam Austin, Carlo Barbieri, Michael Hass, Kurt Snover, Jean-Marc
Sparenberg, Lukas Theussel, and Brian Fields.

\bibliography{s17}

\end{document}